\documentclass[%
 reprint,
 amsmath,amssymb,
 aps,
]{revtex4-1}

\usepackage{graphicx}% Include figure files
\usepackage{dcolumn}% Align table columns on decimal point
\usepackage{bm}% bold math
\usepackage{hyperref}% add hypertext capabilities
\hypersetup{
     colorlinks   = true,
     citecolor    = blue,
     urlcolor     = blue,
     linkcolor    = blue
}
\usepackage[mathlines]{lineno}% Enable numbering of text and display math
\usepackage[all]{hypcap}

\begin{document}

\preprint{APS/123-QED}

\title{Initializing anisotropic and unstable electron velocity distributions \\
needed for investigating plasma kinetic instabilities}

\author{C.-K. Huang }
\affiliation{University of California Los Angeles Department of Electrical Engineering,
Los Angeles, California 90095, USA}
\author{C.-J. Zhang}
\affiliation{University of California Los Angeles Department of Electrical Engineering,
Los Angeles, California 90095, USA}
\author{K. A. Marsh}
\affiliation{University of California Los Angeles Department of Electrical Engineering,
Los Angeles, California 90095, USA}
\author{C. E. Clayton}
\affiliation{University of California Los Angeles Department of Electrical Engineering,
Los Angeles, California 90095, USA}
\author{C. Joshi}
\affiliation{University of California Los Angeles Department of Electrical Engineering,
Los Angeles, California 90095, USA}

\date{\today}

\begin{abstract}
Plasmas with anisotropic electron velocity distribution functions
are needed for the controlled study of kinetic plasma instabilities
in the laboratory. We demonstrate that such plasma can be produced
using ultrashort laser pulses via optical-field ionization (OFI).
We experimentally show this control by using Thomson scattering as
a diagnostic to probe the characteristic electron velocity distributions
using linearly and circularly polarized laser pulses to ionize helium.
Furthermore the He plasma produced by a circularly polarized light
pulse exhibits the onset of the electron streaming instability within
300 fs of ionization, demonstrating applicability of OFI generated
plasmas for studying the kinetic theory regime of plasma physics.

\end{abstract}


\maketitle

The theoretical foundation of plasma physics has a conceptual hierarchy:
exact microscopic or single particle description, kinetic theory and
fluid theory \cite{2016_Book_FFChen}. There are important physical
problems where the complete microscopic description is impractical
while the fluid model is inadequate. In such cases the plasma is described
in terms of one or more velocity distribution functions- this is the
basis of kinetic theory of plasmas \cite{1983_InProc_Davidson}. Experimental
verification of these kinetic effects is predicated upon the ability
to control or know the velocity distribution functions of the plasma
species. For instance temporal evolution of kinetic phenomena such
as plasma wave generation by inverse Landau damping \cite{1995_Book_Goldston}
and instabilities such as the streaming \cite{1992_Book_Stix_WavesInPlasmas},
electron filamentation \cite{1959_PoP_Fried} and Weibel \cite{1959_PRL_Weibel}
could be quantitatively compared with theory if suitable electron
velocity distribution functions (EVDF) could be initialized in a plasma.
Aside from their fundamental interest, these kinetic effects are encountered
in space plasmas \cite{1997_Book_Treumann_SpacePlasmaPhysics}, fast
ignition fusion \cite{2005_PRL_Mendonca}, high-energy colliders \cite{2000_PRL_Ohmi},
neutrino-plasma interactions \cite{1996_PLA_Bingham} and recombination
X-ray lasers \cite{1990_IEEE-QE_Burnett}. With the advent of femtosecond
lasers it has become possible to manipulate the EVDF by optical field
ionization (OFI) of atoms or molecules. Specifically by using an appropriate
combination of laser wavelength(s), intensity profile, polarization,
direction of propagation and ionization state of gases/molecules one
can create plasmas with known EVDF. In this article, we experimentally
demonstrate two examples of such nonthermal and anisotropic distribution
functions by ionizing both electrons of He using fs-class linearly
and circularly polarized laser pulses and show evidence for the electron
streaming instability within 300 fs after the formation of the plasma. 

\begin{figure}[t]
\includegraphics[width=1\columnwidth]{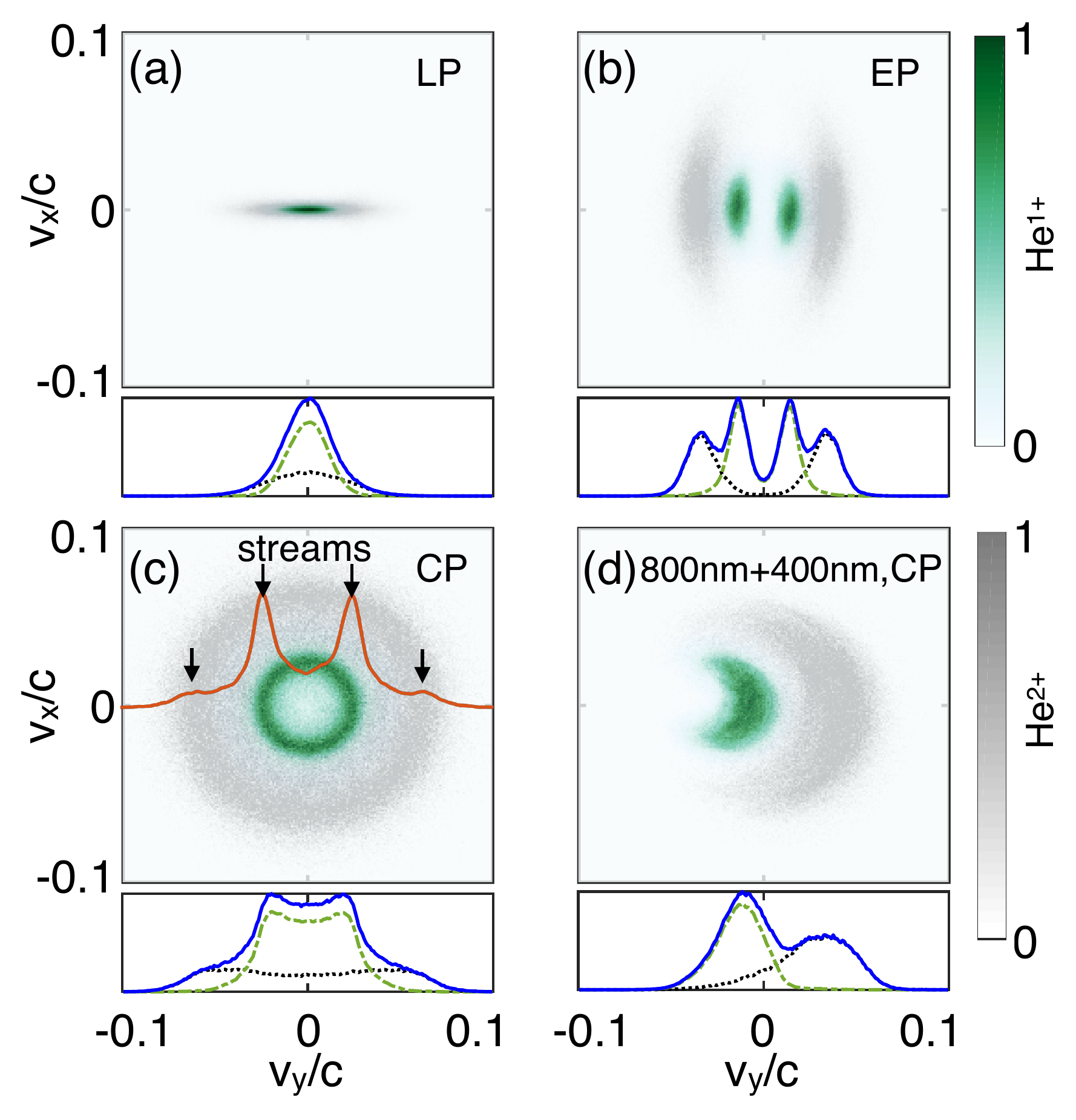}

\caption{(a)-(c) Examples of simulated electron velocity distributions using
OSIRIS of He plasmas produced by 50-fs, 800 nm laser pulses with peak
intensity of $1.6\times10^{17}$ W/cm$^{2}$ and different polarizations
(linear, elliptical, circular respectively). (d) circular, 800 nm,
$3\times10^{16}$ W/cm$^{2}$ + circular, 400 nm, $1\times10^{16}$
W/cm$^{2}$ with the same initial phase. The red curve in (c) shows
the lineout at $v_{x}=0$ showing four streams. Also shown below each
$v_{x}-v_{y}$ image is the $v_{y}$ distribution (the sum of the
number of particles at each $v_{y}$) for He$^{1+}$ electrons (dashed
green curve), He$^{2+}$ electrons (dotted black curve) and their
sum (solid blue line). 
\label{fig1}}
\end{figure}

Optical-field ionization of gases becomes dominant over multi-photon
ionization when the Keldysh parameter is in the tunnel ionization
regime, i.e. $\gamma=\left(U_{i}/2U_{p}\right)^{1/2}\ll1$ where $U_{i}$
is the ionization potential and $U_{p}$ is the ponderomotive potential
of the laser \cite{1964_ZETF_Keldysh}. The energy and the direction
of the ionized electron in OFI depends upon the details of the laser
pulse(s) and the ionization state of the gas 
\cite{1989_PRL_Corkum,2014_PRL_Zhang_OFI-two-color,2014_PRL_Dimitrovski,2015_PRA_Mancuso}.
In Fig. \ref{fig1} we show four examples. Generally speaking the
electrons are ejected transverse to the wave vector of the laser pulse
along the direction of its polarization in the non-relativistic limit
($a_{0}\leq1$), producing strongly non-thermal and/or anisotropic
EVDF in the resulting plasma. Here $a_{0}=eA/mc^{2}=eE/m\omega c$
is the normalized laser strength parameter, where $A$ is the vector
potential, $E$ is the laser electric field, and $\omega$ is the
laser frequency. The EVDF of highly charged states produced by relativistic
pulses ($a_{0}\geq1$) in a dense plasma are rather complicated because
they can be affected by numerous other physical effects such as wakefields/parametric
instabilities \cite{2006_SciAm_Joshi,1981_PRL_Joshi}, direct energy
exchange with the laser field \cite{2017_PRL_Shaw} and therefore
will not be considered here. The polarization dependence of OFI produced
electrons has been tested in previous work in either the long-wavelength
\cite{1989_PRL_Corkum} or the barrier suppression limit using very
low-pressure gases \cite{1993_PRL_Mohideen,1997_PRL_McNaught}. Leemans
et al. \cite{1992_PRL_Leemans} showed that it was possible to control
the Raman instability by varying the polarization of a 200 ps CO2
laser produced OFI plasma. Moore et. al. \cite{1999_PRL_Moore} showed that when
intense ($a_{0}\sim O\left(1\right)$), longer laser pulses are used,
the electrons gain additional energy from the ponderomotive potential
of the laser envelope. Glover et. al. \cite{1994_PRL_Glover} used
Thomson scattering diagnostic to fit the scattered light spectrum
from an OFI He plasma produced using a linearly polarized 800 nm pulse
but they did not observe scattering from each of the two ionic species
of He. Thus, no experimental confirmation of the nonthermal and/or
highly anisotropic initial EVDF characteristic of OFI plasmas has
been made to-date even though the kinetic instabilities that follow
the creation of such plasmas have been predicted \cite{2003_Krainov,2006_PoP_Bychenkov,2017_PoP_Vagin}.

% Figure 1

In Fig. \ref{fig1} we show four examples of such EVDF in velocity
space ($v_{x}$ vs $v_{y}$) using the 3D particle-in-cell (PIC) code
OSIRIS \cite{2002_Fonseca_OSIRIS} where the formation of a fully
ionized, dense ($5\times10^{18}$ cm$^{-3}$) He plasma
is modeled using the ADK theory \cite{1986_ZETF_Ammosov}.
In all cases the simulations consider tunnel ionization of electrons
\cite{1964_ZETF_Keldysh} ($\gamma_{\text{He}^{1+}}=0.38$ and 
$\gamma_{\text{He}^{2+}}=0.23$)
and self-consistently include other physical effects such as the ponderomotive
force of the optical pulse, plasma kinetic effects, and wake formation. We shall refer
to the electron that ionizes first as the He$^{1+}$ electron and
second as the He$^{2+}$ electron. The EVDF shown here are just after
the passage of the laser pulse. Here the x-y plane is perpendicular
to the direction of propagation of the laser, z. In these cases the
He$^{1+}$ electrons are ionized early during the risetime of the
laser pulse within a few laser cycles and the He$^{2+}$ electrons
are ionized approximately 10 fs after the first He electron. These
electrons have both transverse (x and y) and longitudinal (z) oscillating
energy of a few eV due to a weak linear wake formed by the laser pulse
\cite{2006_SciAm_Joshi} and the ions are essentially cold in all
directions. We manipulate the
EVDF in Fig. \ref{fig1} by changing the polarization of the laser pulse to ionize He
atoms from linear (a), to elliptical (b) to circular (c, d). Figure
\ref{fig1}(a) shows that the initial electron distribution along
the laser polarization direction (y) in the linear polarization (LP)
case can be well described by a sum of two 1D (near) Maxwellian distributions
with temperatures of 60 eV (He$^{1+}$) and 210 eV (He$^{2+}$) respectively.
In the elliptical polarization case (degree of ellipticity $\alpha=0.5$,
(Fig. \ref{fig1}(b)) the EVDF shows four lobes with the distribution
in x much wider than that in y. Once again the He$^{2+}$ electrons
(gray) are more energetic than He$^{1+}$ electrons. In the circular
polarization (CP) case (Fig. \ref{fig1}(c)), electron distributions
are donut-shaped in the x-y velocity space. In the x-y plane the resulting
electron velocity distribution has 4 radial streams. 
The transverse streams in Fig. \ref{fig1}(b) and the radial streams in Fig. \ref{fig1}(c)
have larger drift velocities than their thermal velocities. Kinetic theory predicts that plasmas with such 
distribution functions are susceptible to developing 
kinetic instabilities \cite{1983_InProc_Davidson}. It is the relative
drift between these streams that gives rise to the electron streaming
instability. The existence of electrons close to zero transverse velocity
suggests that the plasma has already evolved by the end of the laser
pulse, due to collective effects. The overall initial electron distribution in
the circular case is also shown in Fig. \ref{fig1}(c), blue curve.
It indicates a highly non-Maxwellian distribution with much hotter
root-mean-square (rms) temperature of $\sim470$ eV (220 eV and 910
eV for the He$^{1+}$ and the He$^{2+}$ electrons respectively).
In case 1(d) a two frequency CP laser pulse with different intensities
generates a bump-on-tail distribution that would lead to spontaneous
generation of plasma waves via inverse Landau damping. From the above
examples, it is clear that numerous other \textquotedblleft designer\textquotedblright{}
EVDFs are possible by optimization of laser and choice of the ionizing
medium.

As mentioned earlier the measurement of the EVDF is difficult because
plasmas can very quickly develop kinetic instabilities. These collisionless
processes tend to isotropize the initially produced EVDF on a timescale
far shorter than electron-electron collisions alone, estimated to
be tens of ps for typical value of $T_{x,y}/T_{z}$ expected here.
We therefore use the Thomson scattering diagnostic with $\sim90$
fs (FWHM) probe pulses to interrogate the EVDF of the OFI helium plasma
just $\sim300$ fs after ionization is completed. During such a short
time period plasma density evolution due to expansion or recombination
can be neglected.

\begin{figure}[t]
\includegraphics[width=1\columnwidth]{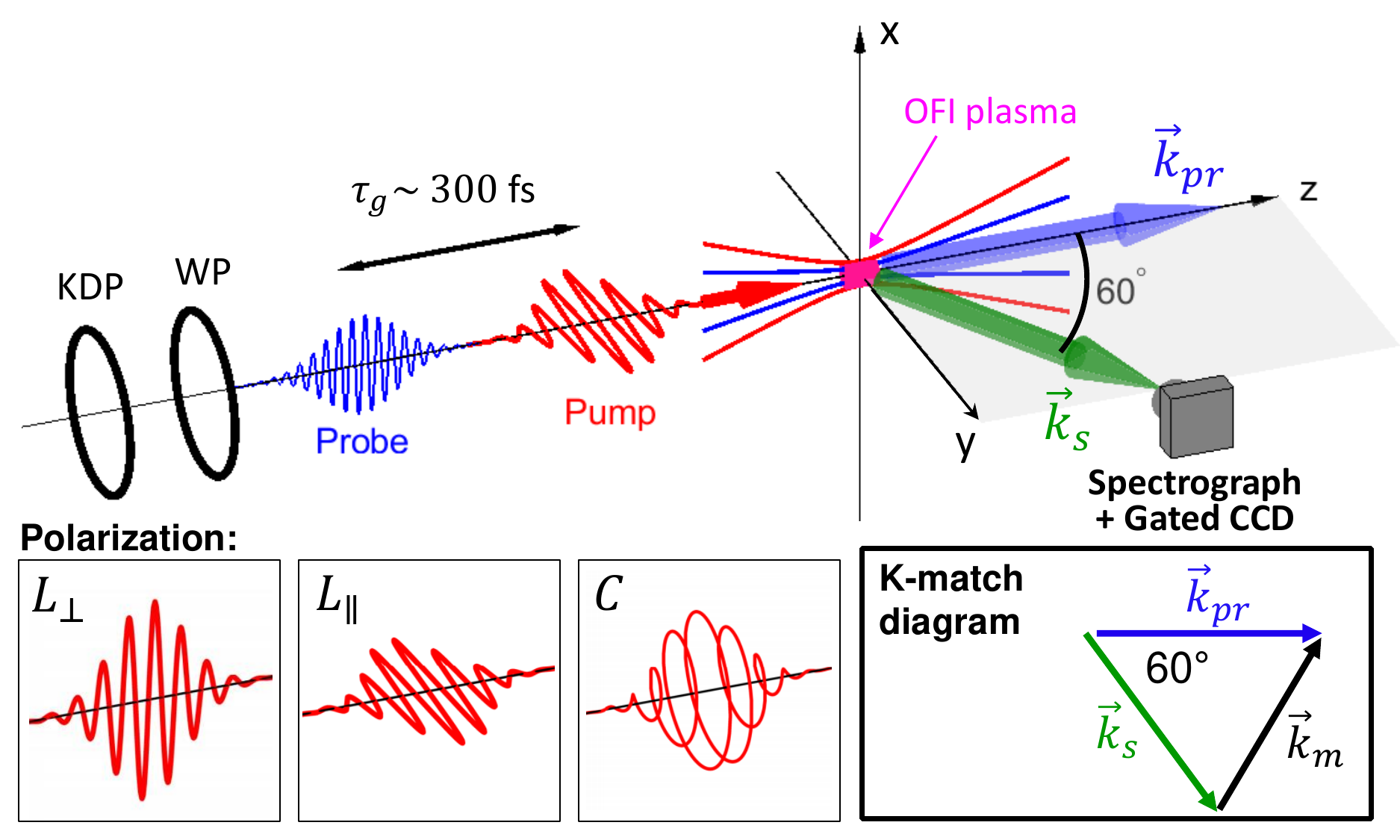}
\caption{Schematic of the experiment. The 800 nm pump beam
generates OFI plasmas that are probed by a collinear 400 nm Thomson
scattering beam using a fixed delay: linear polarization
perpendicular to the scattering plane ($L_{\bot}$), parallel to the
scattering plane ($L_{\parallel}$) and circular polarization ($C$).
Also shown is the k-matching diagram where the vector $\vec{k_{m}}$
is probed in Thomson scattering. KDP: KDP crystal; WP: half-wave plate
for linear polarization or quarter-wave plate for circular polarization.}
\label{fig2}
\end{figure}

The experimental setup is shown schematically in Fig. \ref{fig2}.
The plasma was formed by ionizing a static fill of He gas at various pressures
by focusing a 800 nm, $\sim50$ fs (FWHM) duration laser pulse containing
$\sim10$ mJ energy. The laser was focused by an off-axis parabolic
mirror (OAP) to a spot size $2w_{0}$ of 16 $\mu$m giving a peak
intensity of $\sim1\times10^{17}$ W/cm$^{2}$. The $\sim1$ mJ, $\sim90$
fs (FWHM), 400 nm probe pulse is generated by a 1.5-mm-thick KDP crystal.
The total group delay ($\tau_{g}$) between the pump and the probe
is estimated to be $\sim300$ fs. The probe beam
was focused by the same OAP and focused to a even smaller spot size
within the fully ionized plasma. Thomson scattered light was collected
at $60^{\circ}$ with respect to (w.r.t.) the incident pulse by a
one-to-one imaging system that relays image of the central part of
the plasma to the entrance slit of the spectrograph. The plane containing
the incident probe wave vector ($\vec{k_{pr}}$) and the scattered
light wave vector ($\vec{k_{s}}$) is referred to as the scattering
plane. Two polarization configurations for linearly ($L$) polarized
pump beams are the polarization direction parallel ($L_{\parallel}$)
or perpendicular ($L_{\perp}$) to the scattering plane. The $L_{\parallel}$
($L_{\bot}$) polarization allows us to independently probe the EVDF
essentially along the $v_{y}$ ($v_{x}$) directions as shown in Fig.
\ref{fig1}(a). There is only one configuration for circular polarization
($C$) since the \textquotedblleft double donut\textquotedblright{}
EVDF generated is transversely isotropic (Fig. \ref{fig1}(c)). 

% Figure 2

% Figure 3
\begin{figure}[b]
\includegraphics[width=1\columnwidth]{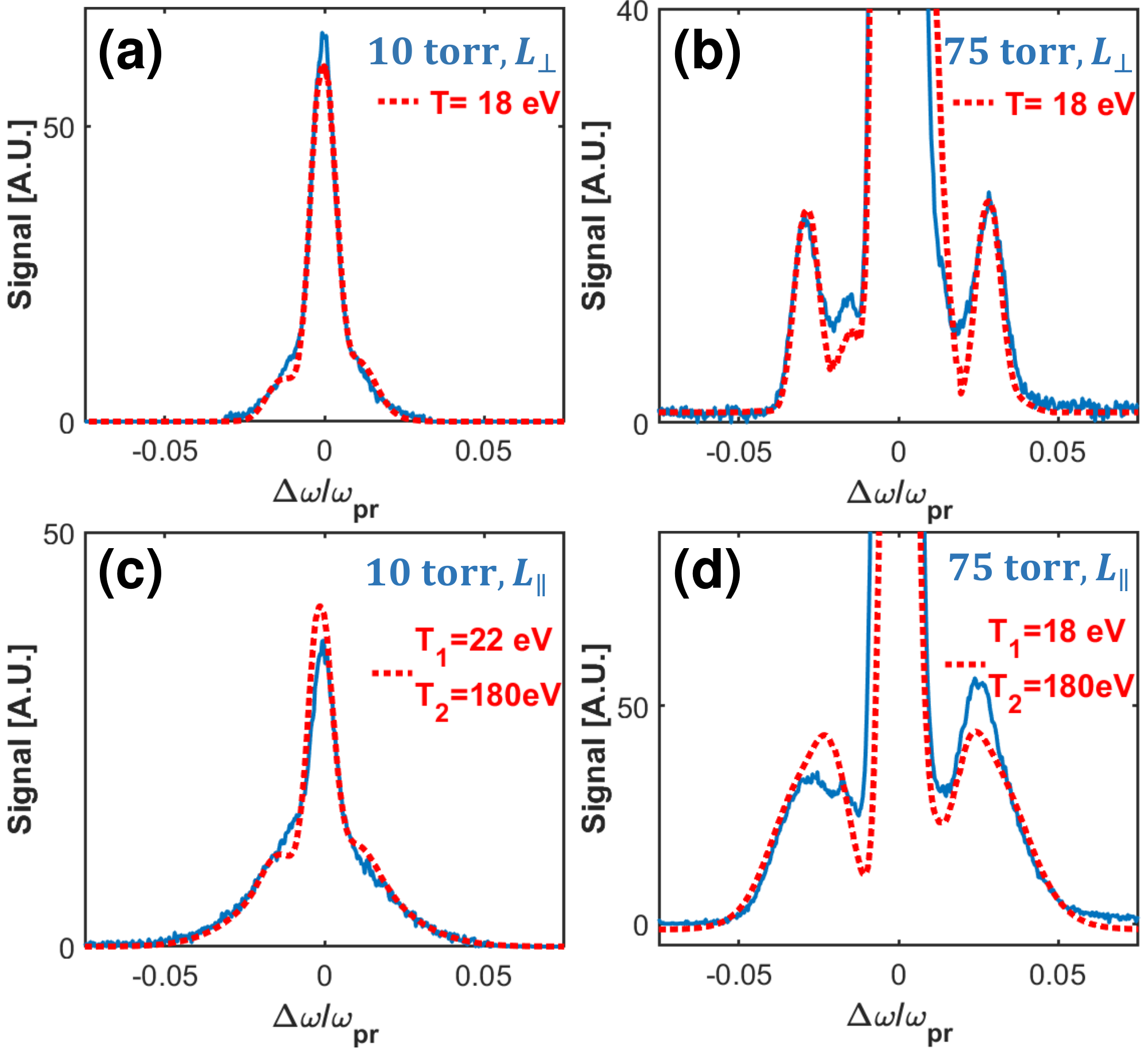}
\caption{Thomson scattering spectra for linear polarization (blue curves- experimental
spectra; dotted red curves- calculated spectra). Polarization direction
is out of the scattering plane for (a) and (b) and parallel to the
scattering plane for (c) and (d). The $L_{\perp}$ cases can be fit
by a single temperature of 18 eV whereas the $L_{\parallel}$ cases
require a two-temperature fit as shown.}
\label{fig3}
\end{figure}

The measured scattered spectra are used to infer the near instantaneous
status of OFI plasmas by comparing them with the Thomson scattering
theory \cite{2011_Book_Froula}. All the data shown in this paper are the average
of 200 consecutive shots to improve the signal-to-noise ratio.
For a non-relativistic, non-magnetized
plasma with an electron distribution function $f_{e}\left(\vec{v}\right)$
and an ion distribution function $f_{i}\left(\vec{v}\right)$, Thomson
scattering spectral power density (SPD) function can be written as

% Equation 1
\begin{equation}
S\left(\vec{k},\omega\right)=\dfrac{2\pi}{k}\left|1-\dfrac{\chi_{e}}{\epsilon}\right|^{2}f_{e}\left(\dfrac{\omega}{k}\right)+
\dfrac{2\pi Z}{k}\left|\dfrac{\chi_{e}}{\epsilon}\right|^{2}f_{i}\left(\dfrac{\omega}{k}\right)\label{eq1}
\end{equation}

\noindent where Z is the atomic number of the atom, $\epsilon=1+\chi_{e}+\chi_{i}$
is the dielectric function, $\chi_{e}$ and $\chi_{i}$ are the electron
and ion susceptibilities. We can apply arbitrary distribution functions
$f_{e}$ and $f_{i}$ to calculate $S\left(\vec{k},\omega\right)$
and get the spectral shape of the Thomson scattered light. 
Due to the broad bandwidth of the probe beam ($\sim3.4$ nm) and the
limiting wavelength resolution ($\sim1$ nm) of the spectrograph,
the ion feature spectrum is not resolved in our experiment and thus
information about the plasma comes from the first term in Eq. (\ref{eq1}).
The $60^{\circ}$ scattering angle determines the measured $\vec{k_{m}}$
in this experiment as depicted in Fig. \ref{fig2}. It should be noted
that the temperatures of the two-Maxwellian distributions in the experiments
are expected to be different than those from the simulations since
we observe the plasma along $\vec{k_{m}}$ which has a $30^{\circ}$
angle with respect to the transverse plane used in simulations. The
observable temperatures, which are evaluated from the projection of
the distribution onto the measured wavevector \cite{2000_Chegotov},
are about 45 and 160 eV for polarization $L_{\parallel}$. 

The scattered light spectra from plasmas produced by LP pump taken
at two fill pressures are shown in Fig. \ref{fig3}. The central spectral
feature at around 400 nm is the ion feature which is not frequency
resolved in this experiment. The frequency shift of the \textquotedblleft electron
feature\textquotedblright , which is associated with collective scattering
from electron plasma waves is symmetric on either side of the ion
feature. The red dashed line in each plot is the best fit of the calculated
SPD function $S\left(\omega\right)$. Figure \ref{fig3}(a) and \ref{fig3}(b)
show the spectra where the polarization is perpendicular to the scattering
plane ($L_{\bot}$). We found that a single Maxwellian distribution
with electron temperature of $18\pm2$ eV (room temperature ions)
fits spectra obtained at both low (10 torr) and high (75 torr) pressures. The corresponding
temperature in the perpendicular plane after 300 fs is expected to
be $\sim12$ eV from simulations. Thus there is a reasonable agreement
between the experiment and the simulations. 

The scattering spectra when the linear polarization is in the scattering
plane $L_{\parallel}$ are shown in Fig. \ref{fig3}(c) and \ref{fig3}(d)
also for helium fill pressures of 10 and 75 torr respectively. In
this case, the calculated SPD functions given by a single Maxwellian
distribution (not shown) do not fit with the experimental spectra.
The data were therefore fitted by taking a two-temperature distribution into consideration. 
Substituting $f_{e}=0.5f_{e,T_{1}}+0.5f_{e,T_{2}}$ into Eq. (\ref{eq1})
where $T_{1}$ and $T_{2}$ are fitting parameters ($T_{2}>T_{1}$) while keeping the
ions as a fixed ultra-cold component, we get a new set of SPD functions
that describe the scattering spectra for the linear polarization case.
The best fits give $T_{1}=20\pm2$
eV and $T_{2}=180\pm20$ eV. The agreement here with the simulations
is again reasonable. We can see that the theoretical plots shown Fig.
\ref{fig3}(c) and \ref{fig3}(d) fit less well than those for Fig
\ref{fig3}(a) and \ref{fig3}(b) both taken at the same pressure
but in the orthogonal plane. 

% Figure 4
\begin{figure}[t]
\includegraphics[width=1\columnwidth]{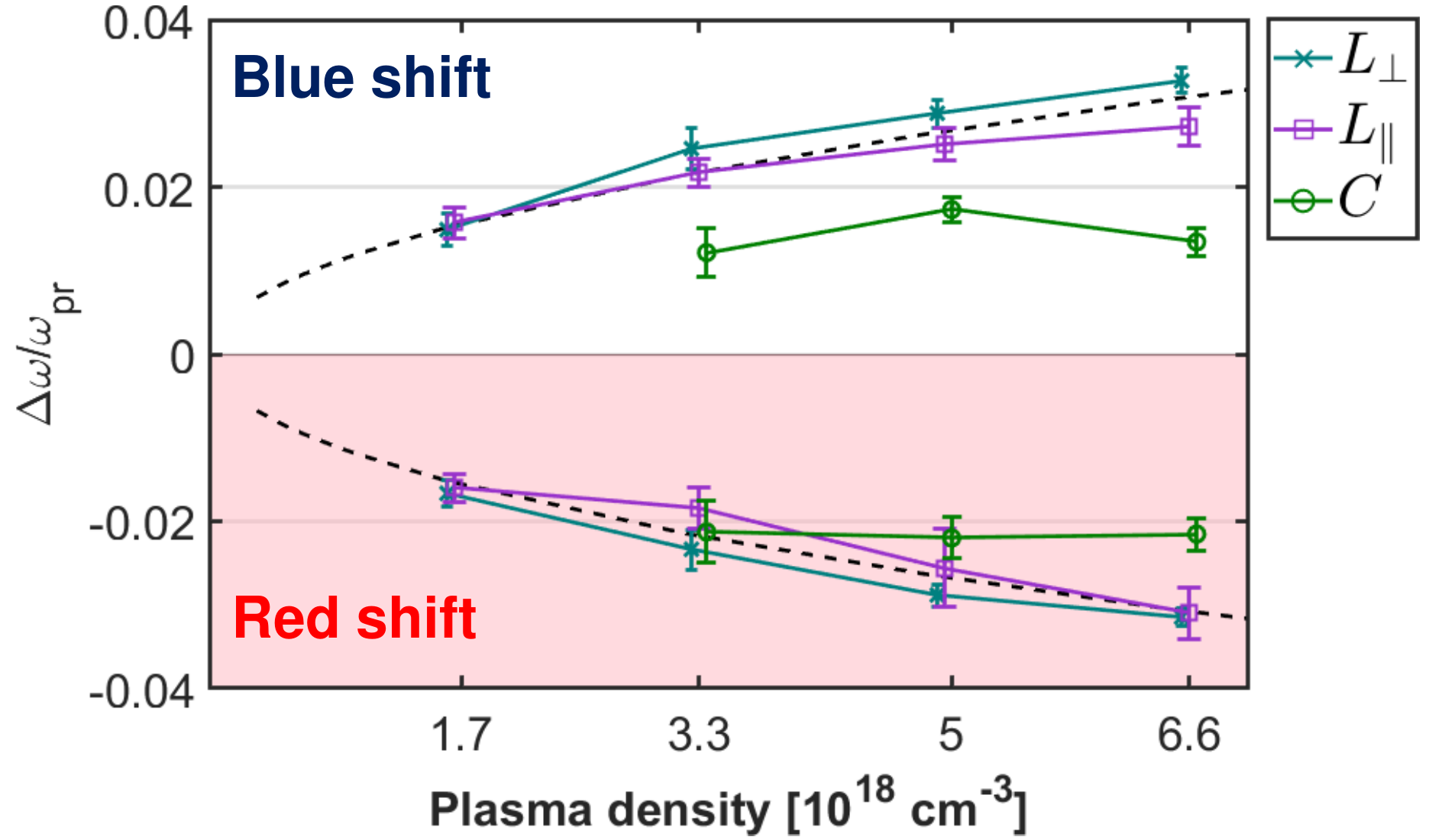}
\caption{The measured spectral peak shifts of the electron feature for different
plasma densities and different laser polarization ($L_{\bot}$, $L_{\parallel}$,
$C$). The error bars show the standard deviation of the shifts for
100 shots. 
The plasma densities plotted correspond to fully ionized He at 25, 50, 75, and 100 torr. 
The dashed lines show the variation of frequency shift
equal to the plasma frequency, $\Delta\omega=\omega_{pe}\left(n_{e}\right)$.}
\label{fig4}
\end{figure}

The frequency shift of the electron feature in the collective scattering
regime should increase as the Langmuir wave frequency, $\omega_{pe}$.
Figure \ref{fig4} shows the measured spectral peak shifts for various
plasma densities for different polarization configurations. For both
$L_{\perp}$ and $L_{\parallel}$, the shifts of their sideband peaks
both increase with densities as expected. This is clearly not the
case in the case of circular polarization which is also shown. The
frequency shift of the electron feature for the CP case was almost
independent of the plasma density, which is indicative of some other
collective phenomena being dominant collective scattering mechanism
than the usual Langmuir waves.

% Figure 5
\begin{figure}[t]
\includegraphics[width=1\columnwidth]{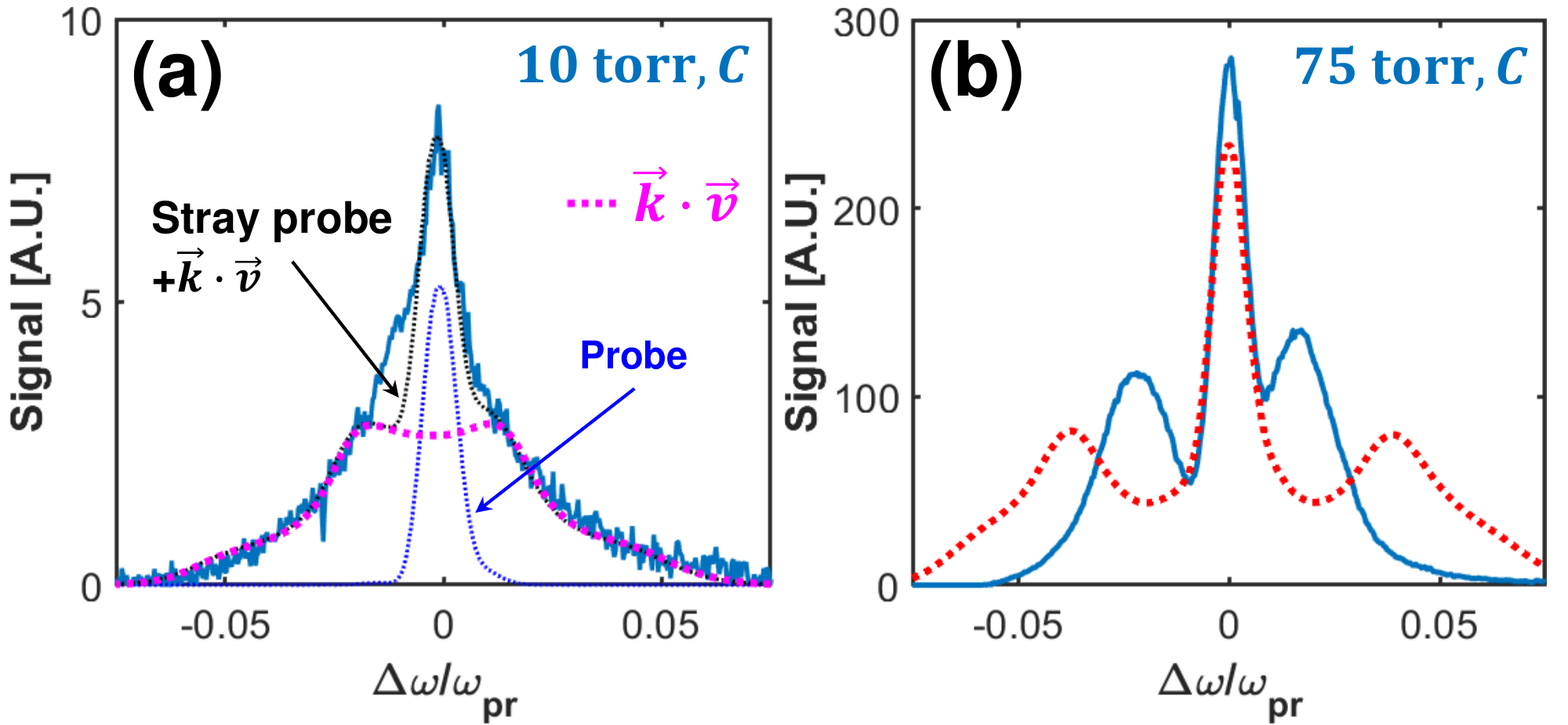}
\caption{Thomson scattering spectra for circular polarization averaged over
200 shots: (a) The measured spectrum at 10 torr He pressure and a
fit that is the sum of the Doppler shifted spectrum (dotted pink curve)
expected from the electron distribution shown in Fig. 1(c) and stray
light spectrum of the probe beam (dotted blue curve). (b) The measured
Thomson scattered spectrum at 75 torr (blue curve) and the calculated
spectrum (dotted red curve) using a distribution with two pairs of
drifting Maxwellian counter streams (drift velocities of $\pm0.015c$
and $\pm0.046c$, widths of 87 and 79 eV, and a density ratio of $\sim4:1$)
deduced from the EVDF shown in Fig. \ref{fig1}(c).
\label{fig5}}
\end{figure}

The Thomson scattered spectra for the CP pump pulses are shown in
Fig. \ref{fig5}. Recall that the electrons in this case have higher
average kinetic energy than those with LP and the EVDF deviate greatly
from Maxwellian. Our fitting attempt using Eq. (\ref{eq1}) failed
with either one-temperature or two-temperature Maxwellian distributions
as expected. 
At low enough plasma densities collective effects are not important
and one expects photons to be Doppler up or down shifted because of
the individual electron motion irrespective of the shape of the EVDF.
We found that it is possible to fit the experimental spectrum taken
at this low plasma density using the distribution function observed
in the simulation as shown in Fig. \ref{fig1}(c). 
The simulated EVDF fits to the wings of the total
spectrum with a plasma density of $6.6\times10^{17}$ cm$^{-3}$ used in the experiment.
When the spectrum of the stray probe photons is also taken into account the overall
Doppler shifted plus the stray photon spectrum fits the experimentally
measured spectrum extremely well.
This excellent fit confirms that the EVDF in the CP case has four streams in the
radial direction as shown by the lineout in Fig. \ref{fig1}(c). 
For the higher density case (Fig.
\ref{fig5}(b)) two distinct spectral \textquotedblleft electron\textquotedblright{}
peaks with asymmetric shifts appeared. Their frequency shifts were
both $\leq\omega_{pe}$  and independent
of the plasma density as was not the case with LP shown in Fig. \ref{fig4}.
This is expected if the scattering is from the streaming instability
where the spectral shift depends on the streams' drift velocity, 
$\Delta\omega\sim\vec{k_{m}}\cdot\vec{v_{d}}$
where $\vec{v_{d}}$ is the relative drift velocity between electron
streams which is independent of plasma density. Substituting the observed
spectral shift of two satellites we obtain $|\vec{v_{d}}|$ equal
to $\left(0.02\pm0.002\right)c$ (blue) and $\left(0.025\pm0.005\right)c$
(red) respectively. Since $\vec{k_{m}}=\vec{k_{r}}+\vec{k_{z}}$ we
are actually observing the oblique electron streaming instability.
This is confirmed in the OSIRIS simulations. The streaming instability
onset occurs in the the x-y plane as expected but it quickly spreads
in all three dimensions this in turn leads to the onset of the electron
streaming instability in an oblique direction \cite{2004_PRE_Bret}
that we observe here. This is the first laboratory observation of
the electron streaming instability because of the anisotropy of the
EVDF of the plasma electrons to our knowledge. The reason why one
can measure the density dependence of the plasma frequency using Thomson
scattering in the LP case is that the onset of the two-stream instability
happens almost 1 ps later when LP is used compared to when CP laser
pulse is used. This is because the ionization process itself produces
relative electron streaming in the CP case whereas the fastest He$^{2+}$
electrons have to bounce off the plasma sheath to begin streaming
in LP case \cite{2018_PRL_Zhang}.

In conclusion, we have demonstrated that OFI may be a method for controlling
the initial EVDF in plasmas. We have used Thomson scattering diagnostic
to probe two such EVDF within 300 fs of their initialization by OFI
in He plasmas using different polarization configurations. The scattered
light spectra are consistent with the expected anisotropic distributions.
Until they are isotropized and thermalized such plasmas cannot be
described by the fluid theory and thus present a new platform for
studying kinetic effects and instabilities in laboratory plasmas.

We thank W.B. Mori for useful discussions. This work was supported
by DOE grant DE-SC0010064, NSF grant 1734315, AFOSR grant FA9550-16-1-0139
and ONR MURI award N00014-17-1-2705.

\bibliographystyle{apsrev4-1}
\bibliography{Bib_ucla_ckhuang_v2}

\end{document}